# Measuring the Impact of Technical Debt on Development Effort in Software Projects

Kartik Gupta
*kartikgupta@outlook.com*

*Abstract*—Technical debt refers to the trade-offs between code quality and faster delivery, impacting future development with increased complexity, bugs, and costs. This study empirically analyzes the additional work effort caused by technical debt in software projects, focusing on feature implementations. I explore how delaying technical debt repayment through refactoring influences long-term work effort. Using data from open-source and enterprise projects, I correlate technical debt with practical work effort, drawing from issue trackers and version control systems. Our goal is to provide a framework for managing technical debt, aiding developers, project managers, and stakeholders in understanding and mitigating its impact on productivity and costs.

## I. INTRODUCTION

first proposed the metaphor of technical debt citeCunningham1993, in reference to the temporary sacrifice of code quality in exchange for faster delivery of features. Thus, incurring technical debt might be useful to achieve immediate deadlines but may simultaneously have negative consequences for the future work, such as software complexity growth, the prevalence of bugs and defects, decreased team productivity and augmented work effort, leading to increased costs of development, infrastructure, and management. The impact of this technical debt may affect the daily work of developers, project managers, product owners and other business stakeholders.

Although initially intended as a metaphor, considerable research has since been undertaken into the identification [1], measurement [2], and management [3] of technical debt. Research has also expanded the scope of the term, such that technical debt spans not just activities related to code implementation but the entire software development environment [4]. The research community identified multiple types of debts [5], each with its advantages and consequences if not handled accordingly. Although industry practitioners are aware of its presence [3] [1], there is no standard way of measuring the current and future impact of technical debt on the development and costs of the team. Additionally, due to a lack of vocabulary and the complexities of the phenomenon, developers find it difficult to convey their concerns to project stakeholders [6].

For example, code smells are an example of technical debt items that may increase future development effort and maintenance costs [7]. Industry case studies have shown that such code violations are more likely to be addressed if made visible by developers within the project's issue tracker [1]. However, from the perspective of a project manager, it is essential to understand how much effort the team puts into technical debt reduction activities and if there is an associated business value. A key concern, therefore, is the ability to measure and make rationale cost-benefit analyses of the impact of technical debt on productivity within a project, capturing both the principal debt incurred and consequent interest as poor code quality accumulates.

The primary research objective of our work is to identify the accumulated additional work effort caused by the presence of technical debt in a code base through an empirical analysis of the work effort of feature implementations within software projects. I intend to determine whether a decision to delay repaying technical debt through refactoring work creates additional work effort in the longer term to repay that debt. I.e., does technical debt really incur interest to the principal, and if so, in what way? An additional objective is to understand how the measure of technical debt varies with software evolution and what types of features incur or reduce its presence. The research questions for our work, organised using the Goal-Question-Metric approach [8], are as follows:

- **RQ1**. Can technical debt be measured in the context of work Effort?
- **RQ1.1**. How does the work effort cost of feature implementation vary with the magnitude of technical debt?
- **RQ1.2**. Does technical debt incur interest in terms of increasing work effort cost of feature implementation as removal of technical debt items are delayed?
- **RQ2**: What are the development patterns surrounding the feature lifecycle?
  - **RQ2.1**: At what checkpoints in feature development is technical debt reduction (refactoring) most prominent?
  - **RQ2.2**: What type of work items incur the most technical debt?

I intend to adopt an empirical approach to the identification of the impact of technical debt on the implementation of future features in existing software code bases. To do this, I will calculate the level of **technical debt** in the system for each change-set recorded in a project's version control repository. I will also record and calculate the **practical work effort** spent on feature implementation linking data from a project's issue tracker and version control repository, thus estimating the effort a developer spent on completing a feature. The impact of technical debt on work effort will then be calculated by comparing the practical work effort with both the estimated effort for an issue.

As data candidates, the study will use both open-source and enterprise projects in order to understand the differences between the two worlds in the context of technical debt work effort. A large Fortune 500 institution has agreed to collaborate with us by granting access to their software development environment data. The candidates will be selected according to various criteria, such as feature development model, type of version control system, availability of estimated work effort metadata, and suitability of code quality tools.

Our long-term goal is to provide a decision-making framework for the management of technical debt in software projects. This framework may provide essential historical information that might be useful for all roles within the software development environment. For example, individual developers could find out how much extra effort was spent in areas with a high number of technical debt items. Architects could identify architectural bottlenecks quickly, while project managers would assign work based on empirical evidence gathered from past work efforts. As of January 2018, the study is in its initial stages.

The rest of this paper is structured as follows. In Section II, I contrast our study with others and how I build upon current research. Section III highlights the proposed work followed by its limitations in Section IV and Section V summarises the proposal.

## II. RELATED WORK

A number of studies have investigated the impact of technical debt items on the costs of software projects. This section reviews this work and relates the existing state of the art to our proposed experiment to empirically measure the work effort interest accumulated by technical debt.

Olbrich et al. [9] studied the impact of two code smells, God class and Shotgun Surgery, on change-proneness of class entities and size of changes within two open source systems, Apache Lucene and Apache Xerces. They found that classes containing either smell are more change-prone than other, non-smelly classes. In a similar study, Khomh et al. [10] empirically analysed the impact of 29 code smells on the change-sets of 9 releases of two open-source systems. They confirmed the results of the previous studies that code smells increase the number of changes that software undergoes during its evolution. Additionally, they found that classes containing more than one code smell are more change prone than other classes. Charalampidou et al. [11] introduced a study that assessed the interest probability of code smells, which is the probability of a code smell introducing extra changes in future development. Interest probability was calculated by counting the frequency of each code smell and how it correlated with the change-proneness of the module where it resides. The smells studied were Long Method, Conditional Complexity, and Duplicated Code. The results showed that code duplication had the highest interest probability due to the number of changes required to maintain future development. Additionally, the complexity of the cyclomatic method increased the number of changes. Fontana et al. [12] studied what was the impact of removing code smells on code quality metrics such as cohesion, coupling, and complexity and which smell incurred the most debt. They applied refactoring activities for each smell, and the metrics were re-evaluated. The results showed that refactoring of one code smell might provide benefits for some metric qualities but may negatively impact others.

The authors studied the impact of code smells and their removal on the software quality metrics output by code quality tools. Although their research has increased awareness of technical debt, our proposed work aims to focus on what impact code smells have on feature implementation by identifying code smells using automated tools. Additional studies have looked at quantifying implementation costs in the context of technical debt.

Singh et al. [13] calculated interest payments by monitoring development effort and code comprehension. They monitored time spent by developers in classes with known technical debt items. They implemented a tool within the Integrated Development Environment of developers to gather information on class visits and development session times. The interest was quantified as the difference between time practically spent in classes and the ideal time. However, the study was conducted with the input of only one developer over nine months. Additionally, estimating the perfect time spent on development is a challenging task due to social and personal factors such as level of project knowledge, environment familiarity, and programming language preference.

Gomes et al. [14] studied the correlation between software evolution, defect occurrence, and work effort deviation at the release level. The authors extracted data from documentation sources such as test plans, project plans, weekly reports, project source code, and emails. Using this data, they could derive important team information on change sets, effort, quality, test, and size of the system. They measured extra work effort by subtracting the estimated work time and total practical work time. Although information at the release level offers project managers an idea of work effort deviation, it does not show at a granular level what defects slow down the development of a new feature and where the team should focus their refactoring activities.

The initial study by Gomes et al. [14] provided a good start to identifying work effort deviation. However, it only analyzed major releases of a system and did not provide drill-down information on iterations and code commits on which features and code smells had the most effect on the work effort. This is essential information for developers and project managers to prioritize refactoring activities in work iterations, especially in an Agile environment where responding to change is critical [15].

## III. EXPERIMENTAL DESIGN

This study focuses on finding the relation between technical debt and work effort. This will yield more information for developers on time management, project managers on refactoring investment, and stakeholders on understanding technical concepts concerning development costs.

The approach of this study will consist of the following steps:

1) *Identify appropriate data candidates for this study*.

   These are potential software projects that are suitable for study. At best, they should have multiple developers contributing to the project, a medium to large codebase with a good amount of historical data, and an associated issue-tracking software. Ideally, the team would have integrated a continuous code quality tool for tracking code smells throughout software evolution that would help with quick and automatic identification of present and historical code issues.

   The study would also benefit from a mixture of open-source and enterprise software to contrast the differences between the two environments and possibly arrive at general conclusions that may contribute to both worlds.

2) *Identify suitable work items from issue tracker*.

   In this step, the purpose is to understand significant events in the evolution of the data candidates. Ideally, events were tracked in the form of tickets with attached metadata, such as:

   - *Priority* will give a sense of importance to the work item. Finding development patterns based on this field will be unique. Will developers take more time to Design and implement an important work item? Alternatively, are they pressured to deliver and thus introduce more debt in the system?
   - *Estimated Work Effort* is one of the most important fields for calculating extra work effort. This provides the theoretical work time on which to compare the practical work time measured in this experiment. It may be in the form of work hours or a story points.
   - The opening and closing *timestamps* of a ticket may offer valuable information for measuring the practical work effort in hours, if developers change the status of tickets as development progresses.

   Unfortunately, this information is not always available. The most important field is the estimated work effort value. Without it, the extra work is difficult to identify. To mitigate this issue, I will not select feature tickets that do not contain this field.

3) *Identify version control checkpoints*.

   Completed work items can be tracked in the version control repository of the project. Identifying checkpoints will aid in understanding the amount of effort put into the work item by the team and how it diverges from the initial estimation.

   Ideally, the team should have links between revisions of the codebase and the work item in the issue tracker. This would make it easier to find the associated checkpoints.

4) *Measure the amount of work effort for each work item*.

   The purpose is to understand how many practical changes a work item has induced over its lifetime. This could be done in two ways: at the code and issue tracker levels.

   At the code level, it is possible to understand the level of work effort involved by aggregating the number of changes a work item has suffered. A change set consists of the number of lines of code added, deleted, and modified. Granularity can be at the pull request, commit, class, and method levels. Version control systems such as Git provide features for retrieving change-sets between revisions.

   However, identifying work effort from change-sets is a challenging task. It is difficult to quantify in working hours since many changes may be generated automatically by modern refactoring tools in the integrated development environment. An alternative solution is to compare the timestamps between the first and the last commit. The temporal difference might provide a practical estimate of the work effort to resolve the issue. Unfortunately, this case only works when a developer works on a single issue at a time.

   At the ticket level, one can understand the amount of work effort realized by a team member. Ideally, the team forces developers to log the time spent designing and implementing a feature. However, that is not always the case. Alternatively, it would be interesting to retrieve the timestamp of ticket events, such as the opening and closing of an issue. Unfortunately, this might not give an approximate time of work since:

   - The team does not respect the opening and closing of a ticket time according to their development patterns. For example, a developer might start work on an issue before marking it as "In Progress" and thus introduce a margin of error.
   - Tickets might remain open for a long period, while features are implemented relatively quickly.
   - There are differences between enterprise and open-source software. For example, developers might work in the timeframe of 9 AM to 6 PM in an enterprise while in open-source, they are free to work at any time of the day. For example, in an extreme case, a developer marks a ticket as "In Progress" before the end of the working day, and resume working the following morning. In this case, our estimation of approximately 15 hours of development time for this work item would be incorrect.

   For this study, the code-level technique will be implemented. The work effort from issue tracking will be implemented in a parallel study. It will be interesting to gather results from both methods and see how they correlate. Additionally, the two result sets may complement one another and provide an overall effort metric.

5) *Measure technical debt items*.

   The scope of this step is to identify code violations within change sets. For each work item implemented, a set of associated modules, classes, and methods will be affected. Historical code smells can be identified and tracked within the evolution of change sets using a continuous code quality tool.

Ideally, the team would have the code quality tool integrated into their continuous integration environment. If so, then historical code smell data could be leveraged by retrieving it through an API. Alternatively, a code quality tool will be used to analyze the version control checkpoints identified and find code smells that may have an impact on change-sets of a feature.

6) *Analysis and discussion of results.*

The three data sources can be linked together once all steps are fulfilled. Extra work can be correlated to issue tracking information and code quality at the time of development. Technical debt can be classified by type, priority, code smells, and assignee.

Unfortunately, the proposed work is not as straightforward. There are many complexities of the work environment, which cannot be considered from the three data sources. Therefore, I will make some assumptions to simplify the process:

- The team uses Git for version control and follows the "Pull Request" model for implementing changes.
- Only one developer is assigned to an issue.
- A developer works on maximum one issue at a time.
- The team uses an issue tracker consistently. Developers change the status of the ticket according to the progress of their current development . For example, if a team member got started on a work item, she would set the ticket status to "In Progress". Respectively, she would mark the ticket as "Closed" if development ceased her changes were reviewed and integrated into the main branch.
- The team estimates the theoretical amount of work necessary to implement a feature or fix a bug. This measure is attached to each work item selected.
- Development time is between 9 AM and 6 PM UTC. Only work items with opening and closing status between these two values are taken into consideration. For example, if a developer starts work on a feature on Monday 5 PM and finishes it by Tuesday 1 PM, then development time will be considered to be 4 hours (1 hour Monday, 3 hours Tuesday). This assumption will simplify the calculation of reasonable work hours for work items.

These assumptions will reduce the amount of real-world complexity in the experiment and allow for simple validation of the results. For enterprise projects, on-premise interviews with teams may help validate the results.

## IV. LIMITATIONS

The experimental design reported in this paper is a work in progress, although I already recognize the limitations of the approach. For example, there are ten types of debt, as identified in a mapping study by Li et al. [5]. In this work, I only aim to examine code and design debt, as they are the most prominent and that developers and project managers have to deal with on a daily basis. The primary challenge is measuring the implementation interest that accrues over the lifetime of a project. Additionally, with the introduction of assumptions to limit the complexity of the study, I have identified the following threats to the validity of our future results:

- Data candidates may not have sufficient information related to the estimation of work effort. I believe this is specifically true for open-source projects implemented by international teams with contributions from many developers.
- Work effort measurement is a difficult challenge and was deemed unmeasurable by Martin Fowler [16]. We consider work effort measurement as the time taken to deliver the requirements, assuming that all requirements of a system in development bring an associated business value. Restricting development time to an interval and discarding multi-developer work items may reduce the data set considerably.
- Code quality tools may not detect potential technical debt items.
- The assumptions made in the previous section may not reflect the real work environment. For instance, developers may be working overtime if under the pressure of a release schedule.
- There are many types and complexities of technical debt that were not included. Requirements, architecture, build, infrastructure, and testing influence the amount of effort needed to finalize a feature.

Although there are risks associated with measurements, care will be taken to consider all possibilities when validating the calculation of measurements. This is especially true in the case of quantifying work effort.

## V. CONCLUSION

To conclude, technical debt is a phenomenon that is difficult to measure accurately and assess potential development and business costs. Therefore, understanding it from the perspective of developers is vital as they are involved in the implementation of new features. Any extra work spent as a result of previously incurred debt increases business costs. If too much debt accrues over the lifetime of a project, the entire project may be brought to a standstill.

As a result, this study will try to understand technical debt from a development work effort perspective. It will possibly shed light on the types of features that take a lot of working hours to complete in correlation with the level of technical debt at the time of the implementation. The first step of the study will identify project candidates from the open-source and enterprise worlds, along with historical feature tickets and their associated implementations. The next step will involve measuring the work effort of these features by analyzing version control and issue-tracking metadata. Subsequently, I will use a code quality tool to identify technical debt items such as code smells with all revisions surrounding each feature implementation. Lastly, I will correlate the data sources gathered and verify whether technical debt items have an impact on work effort feature implementation.

Through our contributions, I hope to help developers discover bottlenecks in productivity, managers to allocate appropriate resources for refactoring activities and business stakeholders to become more aware of development concerns.

## REFERENCES

[1] E. Lim, N. Taksande, and C. Seaman, "A balancing act: What software practitioners have to say about technical debt," *IEEE Software*, vol. 29, no. 6, pp. 22–27, 2012.

[2] F. A. Fontana, R. Roveda, and M. Zanoni, "Technical Debt Indexes Provided by Tools: A Preliminary Discussion," *Proceedings - 2016 IEEE 8th International Workshop on Managing Technical Debt, MTD 2016*, pp. 28–31, 2016.

[3] Z. Codabux and B. Williams, "Managing technical debt: An industrial case study," *2013 4th International Workshop on Managing Technical Debt (MTD)*, pp. 8–15, 2013. [Online]. Available: http://ieeexplore.ieee.org/lpdocs/epic03/wrapper.htm?arnumber=6608672

[4] K. Power, "Understanding the impact of technical debt on the capacity and velocity of teams and organizations: Viewing team and organization capacity as a portfolio of real options," *2013 4th International Workshop on Managing Technical Debt, MTD 2013 - Proceedings*, pp. 28–31, 2013.

[5] Z. Li, P. Avgeriou, and P. Liang, "A systematic mapping study on technical debt and its management," *Journal of Systems and Software*, vol. 101, pp. 193–220, 2015. [Online]. Available: http://dx.doi.org/10.1016/j.jss.2014.12.027

[6] P. Kruchten, R. L. Nord, and I. Ozkaya, "Technical Debt : From Metaphor to Theory and Practice," *IEEE Software*, pp. 18–22, 2012.

[7] M. Fowler, K. Beck, J. Brant, W. Opdyke, and D. Roberts, "Refactoring: Improving the Design of Existing Code," *Xtemp01*, pp. 1–337, 1999.

[8] R. van Solingen, V. Basili, G. Caldiera, and H. D. Rombach, "Goal Question Metric (GQM) Approach," *Encyclopedia of Software Engineering*, vol. 2, pp. 1–10, 2002. [Online]. Available: http://doi.wiley.com/10.1002/0471028959.sof142

[9] S. Olbrich, D. S. Cruzes, V. Basili, and N. Zazworka, "The Evolution and Impact of Code Smells : A Case Study of Two Open Source Systems What are code smells ?" *Proceedings of the 2009 3rd international symposium on empirical software engineering and measurement, IEEE Computer Society*, no. April, pp. 390–400, 2009.

[10] F. Khomh, M. Di Penta, and Y. G. Guéhéneuc, "An exploratory study of the impact of code smells on software change-proneness," *Proceedings - Working Conference on Reverse Engineering, WCRE*, pp. 75–84, 2009.

[11] S. Charalampidou, A. Ampatzoglou, A. Chatzigeorgiou, and P. Avgeriou, "Assessing code smell interest probability: A case study," *ACM International Conference Proceeding Series*, vol. Part F1299, 2017.

[12] F. A. Fontana, V. Ferme, and S. Spinelli, "Investigating the impact of code smells debt on quality code evaluation," *2012 3rd International Workshop on Managing Technical Debt, MTD 2012 - Proceedings*, pp. 15–22, 2012.

[13] V. Singh, W. Snipes, and N. A. Kraft, "A framework for estimating interest on technical debt by monitoring developer activity related to code comprehension," *Proceedings - 2014 6th IEEE International Workshop on Managing Technical Debt, MTD 2014*, pp. 27–30, 2014.

[14] R. Gomes, C. Siebra, G. Tonin, A. Cavalcanti, F. Q. da Silva, A. L. Santos, and R. Marques, "An extraction method to collect data on defects and effort evolution in a constantly modified system," *Proceeding of the 2nd working on Managing technical debt - MTD '11*, p. 27, 2011. [Online]. Available: http://www.scopus.com/inward/record.url?eid=2-s2.0-79960595082&partnerID=tZOtx3y1

[15] "Agile Manifesto," http://agilemanifesto.org/.

[16] M. Fowler, "Cannot Measure Productivity," https://martinfowler.com/bliki/CannotMeasureProductivity.html, 2003.